**Optimized polar-azimuthal orientations for polarized light illumination of different Superconducting Nanowire Single-Photon Detector designs**

Mária Csete,[1, 2*] Áron Sipos,[1] Faraz Najafi,[2] and Karl K. Berggren[2]

[1]*Department of Optics and Quantum Electronics, University of Szeged,*

*Dóm tér 9, Szeged, H-6720, Hungary*

[2]*Research Laboratory of Electronics, Massachusetts Institute of Technology,*

*77 Massachusetts Avenue, Cambridge, MA 02139, USA*

**Abstract**

The optimal orientations are determined for polarized substrate side illumination of three superconducting nanowire single-photon detector (SNSPD) designs: (1) periodic niobium-nitride (NbN) stripes standing in air with dimensions according to conventional SNSPDs, (2) same NbN patterns below ~quarter-wavelength hydrogensilsesquioxane-filled nano-cavity, (3) analogous NbN patterns in HSQ nano-cavity closed by a thin gold reflector. Numerical computation results have shown that the optical response and near-field distribution vary significantly with polar-angle, $\varphi$, and these variations are analogous across all azimuthal-angles, $\gamma$, but are fundamentally different in various device designs. Larger absorptance is available due to p-polarized illumination of NbN patterns in P-structure configuration, while s-polarized illumination results in higher absorptance in S-structure arrangement. As a result of p-polarized illumination a global maximum appears on absorptance of bare NbN pattern at polar angle corresponding to NbN-related ATIR; integration with HSQ nano-cavity results in a global absorptance maximum at polar angle corresponding to TIR at sapphire-air interface; while the highest absorptance is observable at perpendicular incidence on P-structures aligned below gold reflector covered HSQ nano-cavity.

S-polarized light illumination results in a global absorptance maximum at TIR on bare NbN patterns; the highest absorptance is available below HSQ nano-cavity at polar angle corresponding to ATIR phenomenon; while the benefit of gold reflector is large and polar angle independent absorptance.

**1. Introduction**

The standard absorbing structure in superconducting-nanowire single-photon detectors (SNSPD) consists of 4 nm thick niobium-nitride (NbN) stripes having a width of ~30-100 nm [1, 2]. In order to ensure appropriate resistive barrier across the boustrophedonic NbN pattern, wire feature sizes of ~100 nm arrayed in 200 nm periodic structure with 50 % filling factor are ideal [2]. SNSPD devices are extensively used for infrared photon counting, however their detection efficiency is strongly limited optically by losses accompanying reflection from and transmission through the simple geometrical structure.

The optimization of SNSPDs requires the maximization of the NbN pattern's absorptance. Two successful approaches realizing this purpose were described in previous literature. The first approach is based on optical device design development, as the device structure fundamentally limits the available absorptance in all detectors. Different types of integrated patterns were designed to reach larger absorptance in SNSPDs [3-6]. In so-called OC-SNSPD devices the absorbing NbN elements are aligned below an optical cavity [3, 4]. Detection efficiency of 50 % was experimentally observed, when an integrated structure consisting of NbN pattern below an optical cavity and a 120 nm thick gold anti-reflection-coating was illuminated by 1550 nm wavelength light incident perpendicularly from substrate side [3].

More complex integrated device structures, like SNSPDs consisting of noble-metal nano-antenna-arrays were experimentally and theoretically studied too, but only in case of perpendicular incidence. These nano-antenna-array integrated devices make possible to reach 96 % absorptance in case of **E**-field oscillation perpendicular to the gold pattern [5, 6].

The second approach is based on optimization of illumination conditions. The absorptance of thin lossy wires inherently depends on the relative orientation of their pattern with respect to **E**-field oscillation direction [7]. Theoretical studies on front-side perpendicular illumination of NbN patterns embedded into dielectric media confirmed that larger absorptance is available in case of **E**-field oscillation parallel to the NbN wires [8, 9].

Latest results in the literature proved that illumination of SNSPD devices at polar angle corresponding to Total Internal Reflection (TIR) is capable of resulting in 100 % absorptance, when an NbN pattern in S-structure arrangement is illuminated by s-polarized light [10]. It was also shown that p-polarized illumination of the same pattern in P-structure configuration causes zero absorptance at TIR. Here we refer to an arrangement as P-structure configuration, when the incidence plane is parallel to the NbN wire-grating, while S-structure arrangement is the case, when the wires are perpendicular to the plane of incidence [11].

The purpose of our present study was systematic investigation of the simplest SNSPD designs in order to determine the optimal conditions for polarized, substrate side, off-axes illumination in conical-mounting. Inspection of illumination direction dependent nano-optical phenomena made it possible to maximize the absorptance in significantly different arrangements of various system designs. Similar studies were realized for nano-cavity-array integrated optical systems applicable in SNSPDs too [12].

## 2. Theoretical methods

Three types of superconducting-nanowire single-photon detector (SNSPD) designs were theoretically studied: (1) 200 nm periodic pattern of 4 nm thick and 100 nm wide NbN stripes standing in air, which are covered inherently by a ~2 nm thick NbNO$_x$ layer (Figure 1a); (2) the same NbN patterns integrated with continuous 279 nm thick, i.e. ~quarter-wavelength hydrogensilsesquioxane filled nano-optical cavity (Figure 1b); (3) analogous cavity-integrated structures covered by a 60 nm thick continuous gold film acting as a reflector (Figure 1c).

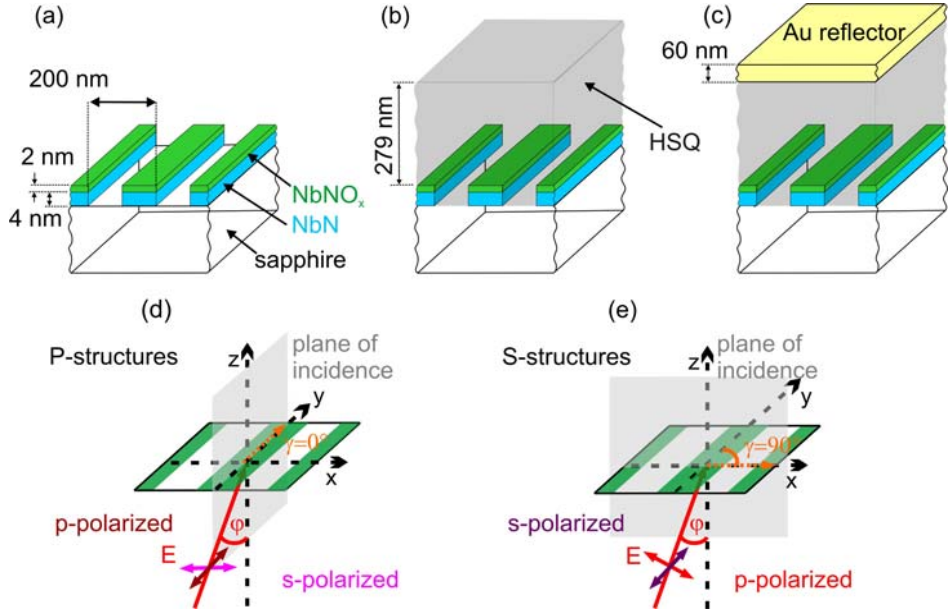

**Fig. 1.** Schematic drawing of the optical systems studied: (a) 200 nm periodic pattern of 4 nm thick and 100 nm wide NbN stripes standing in air, inherently covered by ~2 nm NbNO$_x$ layer; (b) the same structure as in (a) arrayed below HSQ-filled nano-cavity having 279 nm length; (c) the same structure as in (b) covered by a continuous 60 nm thick Au reflector. All patterns are illuminated by $\lambda$ = 1550 nm polarized light from sapphire substrate side, the illumination directions in conical mounting are specified by $\varphi$ polar and $\gamma$ azimuthal angles. The two specific orientations studied in more details are: (d) P-structure configuration ($\gamma$=0°), and (e) S-structure arrangement ($\gamma$=90°). The p- and s-polarized illuminations of both structures are indicated.

Optical system (1) is similar to the pattern studied by Driessen et al. [10] however the layer thickness and the wavelength are different in our present study. Device structure (2) is analogous with the structure studied by Anant et al. [9], but the NbN stripes are illuminated from sapphire substrate side by p- and s-polarized light in our present work. The cavity-based structure (3) is analogous with the integrated pattern studied in our previous works [3, 11], but we investigate off-axes polarized light illumination of NbN stripes with arbitrary **E**-field oscillation direction with respect to their periodic pattern in present study.

*2.1. Finite Element Method to determine the optical response and near-field distribution*

The three dimensional FEM method developed in our previous work based on Radio Frequency module of Comsol Multiphysics software package (COMSOL AB) was applied [11]. Special 3D FEM models were used to determine the effect of polar and azimuthal illumination angles on the optical response and to map the electromagnetic near-field distribution around NbN segments at specific orientations corresponding to extrema on NbN absorptance in optical systems (1-3).

Polarized infrared light beam with $\lambda$=1550 nm wavelength, having a power of $P$=2*10$^{-3}$ W was used to illuminate of NbN patterns from sapphire substrate side. The illumination direction dependent absorptance was determined based on Joule-heating inside NbN and gold segments, while reflectance and transmittance signals were extracted based on power-outflows from the optical systems.

In a coarse dual-angle-dependent study the $\varphi$ polar angle, which is measured relative to the surface normal, and the $\gamma$ azimuthal orientation, specified by the angle between the plane of light incidence and NbN stripes, were varied within the ranges of $\varphi$=[0°, 85°] and $\gamma$=[0°, 90°] with angular resolution of $\Delta\varphi = \Delta\gamma$ = 5° (Figure 1d, e; 2 and 4). The special cases of p- and s-polarized illumination of integrated patterns in P-structure configuration ($\gamma$=0°) and in S-structure arrangement ($\gamma$=90°) were investigated by varying the polar angle with larger $\Delta\varphi$ = 0.05° resolution in intervals surrounding NbN absorptance maxima (Fig. 1d, e). These results were integrated into results of computations performed with lower $\Delta\varphi$ = 1° resolution across the entire $\varphi$=[0°, 85°] polar angle region (Figure 3 and 5).

*2.2. Transfer Matrix Method to determine the optical response*

We performed Transfer Matrix Method calculations on different multilayers aligned on sapphire substrate according to the vertical stacks composing optical systems (1-3) shown in Figure 1a-c [13]. The alternating stacks containing NbN and without NbN are as follows: (1)

NbN-NbNO$_x$ bilayers standing in air and air medium; (2) NbN-NbNO$_x$ bilayers below 279 nm HSQ layer and 285 nm thick HSQ layer; (3) NbN-NbNO$_x$ bilayers below 279 nm HSQ nano-cavity covered by 60 nm Au film and 285 nm HSQ layer covered by 60 nm Au film. The absorptance, reflectance, and transmittance were determined by weighting optical responses of each stack by their corresponding 50 % fill-factor [11]. These results were compared with the results of FEM computations in order to determine the origin of each extrema appearing on the optical response (Figure 3 and 5).

## 3. Results and discussion

The common characteristics of optical responses and near-field distributions are that off-axes illumination results in significant variation as a function of $\varphi$ polar angle, while only a slow and monotonous variation is observable, when $\gamma$ azimuthal angle is tuned from 0° to 90° values (Fig. 2-7). As a result, the extrema are analogous, however the maximal values are different in P-structure configuration and S-structure arrangement (Fig. 2-5).

### *3.1. Dual-angle dependent absorptance resulted by p-polarized light illumination*

In case of p-polarized light illumination the P-structure configuration results in larger available absorptance in all of three studied optical systems (Fig. 2). The absorptance maxima appear in entirely different polar angle intervals in various device designs (Fig. 2, 3; Table I), their origin will be described in sections 3.2.1-3.2.3.

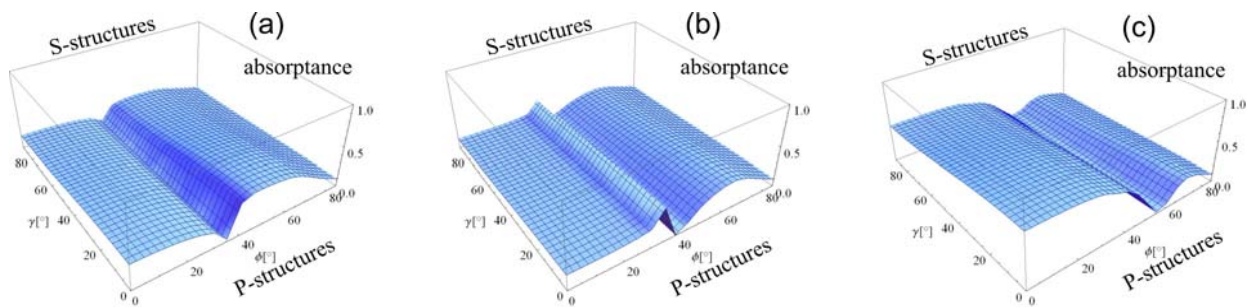

**Fig. 2.** Dual-angle-dependent absorptance due to p-polarized light illumination of NbN patterns (a) standing in air, (b) arrayed below HSQ-filled nano-cavity, (c) aligned below HSQ-filled nano-cavity covered by Au reflector. The absorptance was determined by calculations performed over the $\gamma = [0-90°]$ and $\varphi = [0-85°]$ intervals, with $\Delta\gamma = \Delta\varphi = 5°$ resolution.

Figure 2a indicates that the dual-angle dependent absorptance of NbN pattern standing in air exhibits a narrow minimum at all azimuthal angles, in the interval surrounding the polar angle corresponding to TIR at sapphire-air interface. This absorptance characteristic is very similar to the absorptance signal presented previously for p-polarized illumination of P-structures in reference [10]. When NbN pattern is aligned at the bottom of quarter-wavelength HSQ nano-cavity, the dual-angle dependent absorptance does not indicate this narrow minimum, in contempt of p-polarized light illumination (Figure 2b). On the contrary, this curve is more similar at all azimuthal angles to the polar angle dependent absorptance, which was previously presented for s-polarized illumination of S-structures in reference [10]. The fundamental difference in presence of gold reflector with respect to optical systems (1) and (2) is that the absorptance is more enhanced at perpendicular incidence and throughout small polar angles, rather than at large tilting. This indicates that the application of reflectors makes it possible to reach higher absorptance values in experimentally more easily implementable arrangements in case of p-polarized illumination (Fig. 2c).

***3.2. Polar-angle dependent optical responses resulted by p-polarized light illumination***

The high resolution computations show that sudden changes are observable at the polar angle corresponding to TIR at sapphire-air interface in all of three studied systems (Fig. 3).

*3.2.1 Optical response of NbN pattern standing in air*

Figure 3a shows that P-structures result in larger absorptance than S-structures in case of p-polarized 1550 nm illumination of bare NbN patterns through all polar angles.

At perpendicular incidence the absorptance of P-structures is larger than the absorptance predicted by TMM, and approximately two-times larger than the absorptance observed in S-structure-arrangement [11]. A global minimum is observable on absorptance curve of both P- and S-structures at 34.7° polar angle corresponding to TIR at the sapphire-air interface. This is in accordance with zero absorptance at TIR presented in previous literature [10].

The differences in absorptance values observed in P-structure configuration with respect to reference [10] can be explained by the different NbN layer thicknesses and by the different wavelength of light applied for illumination.

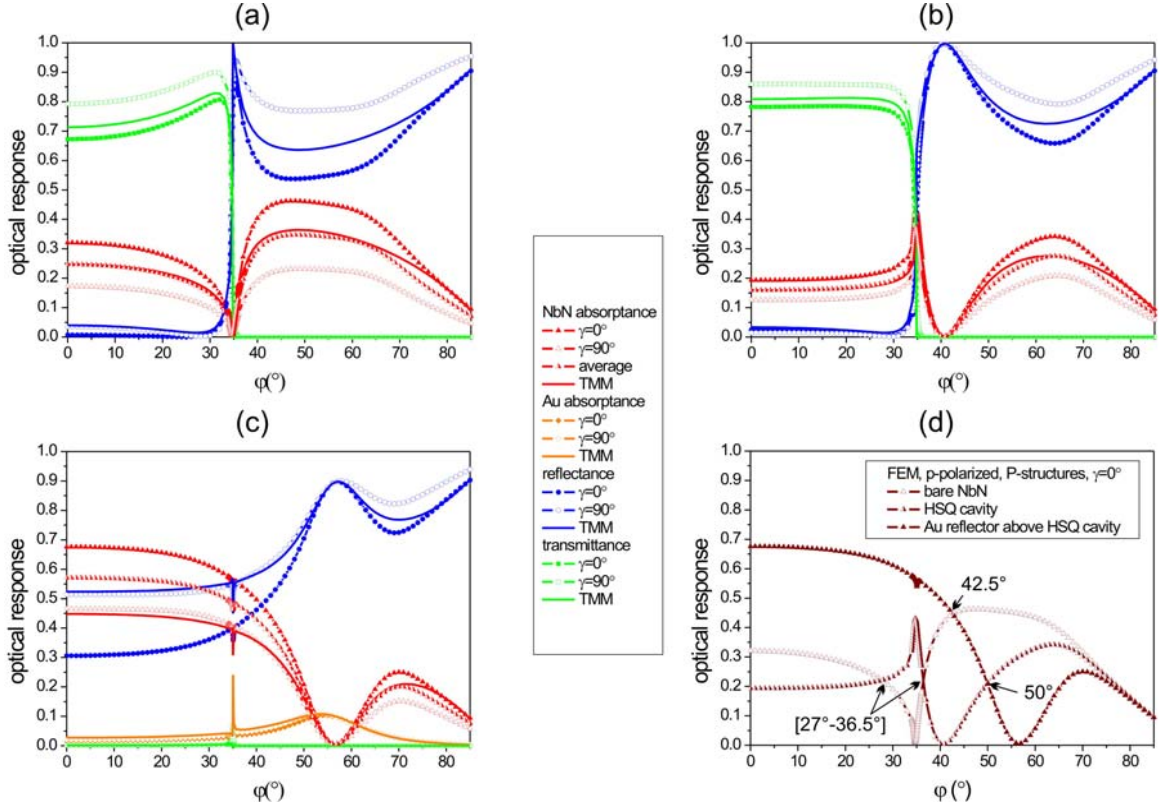

**Fig. 3.** The comparison of optical responses in P-structure configuration ($\gamma$=0°, closed symbols) and in S-structure arrangement ($\gamma$=90°, open symbols) determined by FEM with optical responses determined by TMM (lines) for the case of p-polarized light illumination of NbN patterns (a) standing in air, (b) arrayed below HSQ-filled nano-cavity, (c) aligned below HSQ-filled nano-cavity covered by Au reflector. The results of $\Delta\varphi = 0.05°$ resolution FEM and TMM calculations in [34°, 36°] interval are incorporated into the graphs originating from computation performed with $\Delta\varphi = 1°$ resolution in [0°, 85°] region. (d) Comparison of polar angle dependent absorptance in three optical systems for the case of p-polarized illumination of P-structures.

The global maxima are at ~48° and 49° polar angles, where the absorptance reaches 46.3 % value in P-structure configuration and 23.1 % in S-structure arrangement, due to NbN-related ATIR. Based on these observations the optimal orientation of NbN pattern standing in air is P-structure-configuration tilted to 48° polar angle. However illumination under this orientation is experimentally difficult, as it would require light in-coupling into the system above polar angle corresponding to TIR phenomenon.

The transmittance indicates a cut-off at TIR, while the reflectance exhibits ATIR phenomenon according to presence of lossy NbN pattern at sapphire-air boundary. The reflectance is most strongly frustrated at orientations optimal for NbN absorption.

The average of FEM absorptances equals with the TMM absorptance, while in P-structure configuration FEM absorptance is significantly larger than the TMM absorptance through the entire polar-angle interval. This difference is caused by the inherent limits of TMM [12], which method cannot account for absorption modification occurring, when the projection of the **E**-field oscillation is varied with respect to the lossy stripes [7-9].

*3.2.2. Optical response of NbN pattern below HSQ-filled nano-cavity*

Figure 3b presents that the TIR has a pronounced effect on the absorptance in both P-structure and S-structure arrangements for p-polarized light illumination of HSQ-cavity integrated optical system (2).

In case of perpendicular incidence the absorptance values are smaller than the absorptance in analogous arrangements without HSQ nano-cavity. Then the absorptance increases, and surprisingly reaches a global maximum at 34.8° incidence angle, which is the polar angle corresponding to TIR at sapphire-air interface. At this tilting the enhanced 43.1 % absorptance of P-structures and the 31.3 % absorptance of S-structures are larger than the absorptances at perpendicular incidence on either of bare or HSQ-covered NbN patterns in analogous arrangements. These observations prove that in presence of quarter-wavelength HSQ nano-cavity off-axis illumination may result in ~50 % absorptance in P-structure-configuration at considerably smaller polar angle, than on bare NbN patterns.

In this device design the absorptance indicates a global minimum at 41°, then a local absorptance maximum appears at 64° originating from NbN-related ATIR. At this local maximum the absorptance values are only slightly larger than the absorptance at perpendicular incidence on bare or HSQ-covered NbN patterns in analogous arrangements.

The intermediate values indicate that the pronounced tilting does not have real advantages, as it is hard to realize such large incidence angles experimentally, and the available absorptance is smaller than the global maximum observed on NbN pattern standing in air.

The transmittance indicates again a cut-off at 34.8° polar angle, while the reflectance curve exhibits a Brewster-like minimum at 28°, followed by an inflection point at 34.8° polar angle. The total reflectance is shifted to 41° in presence of HSQ layer, while above this polar angle the reflectance exhibits ATIR characteristics caused by presence of lossy NbN stripes.

The average of FEM signals equals with the TMM absorptance again, and the FEM absorptance in P-structure configuration is only slightly larger, while in S-structure arrangement it is slightly smaller than the TMM absorptance across all polar-angles. The difference between the optical responses in P- and S-structure arrangements is smaller than in case of NbN pattern standing in air, and it is almost negligible at the global maximum and at large polar angles. The TMM absorptance curve indicates an analogous global maximum at 34.8° polar angle, however this method predicts smaller maximal value.

*3.2.3. Optical response of NbN pattern below Au reflector covered HSQ-filled nano-cavity*

Figure 3c indicates that already small polar angle intervals are promising for detection, when NbN pattern is illuminated by p-polarized light in gold reflector covered quarter wavelength HSQ cavity. The highest available absorptance observed at perpendicular incidence is 67.6 %, when the stripes are in P-structure-configuration, which value is approximately 1.5 times larger, than the 46.5 % absorptance observed in case of S-structures [11].

There are narrow local minima close to the tilting corresponding to TIR at sapphire-air interface, where the absorptance is reduced slightly. The 35.05° polar angle position of minima indicates that the origin of these narrow extrema is ATIR phenomenon. A broad global minimum appears at 57°, which is followed by a small local maximum at 70° in both P- and S-structures originating from NbN-related ATIR.

The gold refletor absorptance indicates a narrow local maximum at 35.15° and broad global maxima at 55° and 53° incidence angles in case of P- and S-structures, confirming that the minima in NbN absorptance are in polar angle intervals, where the TIR is frustrated caused by losses in gold, too. These orientations result in surface plasmon mode excitation, since the projection of the light wave vector matches the wave vector of plasmons propagating at air-gold and at HSQ-gold boundary at these polar angles [11].

The transmittance is suppressed in entire polar angle interval, only a slight transmittance is observable at 35.05° on the FEM curves. These little artifacts are caused by near-field signal subtraction in close proximity of the gold reflector supporting SPPs.

The average of the FEM signals does not equal with the TMM optical responses, as the S-structure absorptance approximates it through the entire polar angle interval, while the P-structure absorptance is considerably larger throughout the global minimum. After this extremum there is only a slight difference between the signals predicted by FEM and TMM.

The optical system (3) might be considered as grating-waveguide configuration, which supports polar angle dependent Fabry-Perrot resonances [14]. This explains the pronounced effect of tilting on the optical response.

*3.2.4. Comparison of different systems illuminated by p-polarized light*

The fundamentally different absorptances available in various optical systems are compared in Fig. 3d and Table I for the optimal azimuthal orientation of P-structure configuration in case of p-polarized light illumination.

Figure 3d indicates that the reflector results in the highest NbN absorptance in a wide polar angle interval. Detailed comparison of the absorptance in optical systems (1) and (2) shows that it is possible to enhance the absorptance with respect to bare NbN patterns already without a gold cover layer throughout the [27°-36.5°] polar-angle interval in P-structure configuration.

| Tilt \ Media | Perpendicular | | TIR | | NbN | |
|---|---|---|---|---|---|---|
| | φ (°) | A | φ (°) | A | φ (°) | A |
| air | 0 | 0.32 | 34.7 | 0.01 | 48 | **0.463** |
| HSQ | 0 | 0.193 | 34.75 | **0.43** | 64 | 0.341 |
| Au+HSQ | 0 | **0.676** | 35.05 | 0.539 | 70 | 0.25 |

Table I. Summary of absorptance values observed at perpendicular incidence, determined by TIR and NbN-related ATIR phenomenon on polar-angle dependent absorptance of three different optical systems, when the NbN pattern in P-structure-configuration is illuminated by p-polarized light. The absorptances at optimal orientations are emphasized with characters in bold.

In presence of a reflector the absorptance is larger, than in case of NbN pattern standing in air trough 42.5° incidence angle, while it crosses the absorptance curve of NbN-pattern below HSQ nano-cavity at 50° polar angle.

### *3.3. Dual-angle dependent absorptance resulted by s-polarized light illumination*

In case of s-polarized light illumination the S-structure arrangement results in larger available absorptance in all of the studied optical systems (Fig. 4). The absorptance maxima appear in similar polar angle intervals in different device designs (Table II), their origin will be described in sections 3.4.1-3.4.3.

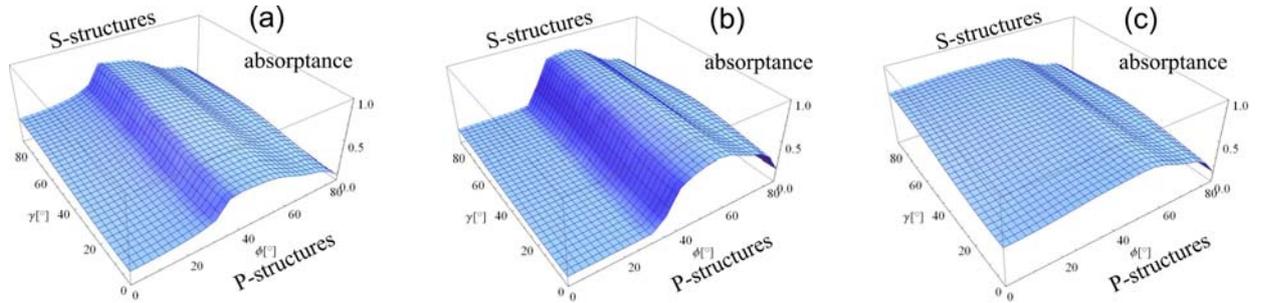

**Fig. 4.** Dual-angle-dependent absorptance due to s-polarized light illumination of NbN patterns (a) standing in air, (b) arrayed below HSQ-filled nano-cavity, (c) aligned below HSQ-filled nano-cavity covered by Au reflector. The absorptance was determined by calculations performed over the $\gamma = [0-90°]$ and $\varphi = [0-85°]$ intervals, with $\Delta\gamma = \Delta\varphi = 5°$ resolution.

Figure 4a indicates that the dual-angle dependent absorptance of NbN pattern standing in air exhibits a global maximum exactly at the polar angle corresponding to TIR at sapphire-air interface at all azimuthal angles.

This is analogous with the absorptance signal presented previously for s-polarized illumination of S-structures in reference [10]. When the NbN pattern is at the bottom of an HSQ nano-cavity, the dual-angle dependent absorptance indicates a broadened maximum at larger polar angles at all azimuthal angles in case of s-polarized light illumination (Figure 4b).

The gold reflector integration results in that the absorptance is uniformly enhanced from perpendicular incidence throughout large polar angles however the maximal values are somewhat smaller in comparison to optical systems (1) and (2). Optical system (3) seems to be the most advantageous also in case of s-polarized illumination, as high absorptance values are available in experimentally easily implementable arrangements (Fig. 4c).

### *3.4. Polar-angle dependent optical responses resulted by s-polarized light illumination*

The high resolution computations show that the maxima surround the polar angle corresponding to TIR at sapphire-air interface in all of three studied systems (Fig. 5). Interestingly, in case of s-polarization both FEM and TMM computations predict smaller differences between absorptances available in different basic systems in comparison to p-polarized light illumination.

The absorptance values observable, when NbN patterns are illuminated in specific configurations by perpendicularly incident s-polarized light, equal with absorptances observed, when p-polarized light illuminates NbN patterns in complement arrangements.

### *3.4.1 Optical response of NbN pattern standing in air*

Figure 5a shows that TIR determines the optical response of NbN pattern standing in air. At perpendicular incidence the S-structures' absorptance is larger than the absorptance predicted by TMM, and approximately two-times larger than the absorptance observed in P-structure-arrangement, i.e. the ratio is reversed in comparison to p-polarized case. The absorptance indicates global maxima of 75 % and 47.9 % at 34.85° and 37.45° polar angles in S- and P-structures. The optimal orientation for s-polarized illumination of NbN pattern standing in air is S-structure arrangement under polar angle corresponding to TIR.

The absorptance curve is similar to Driessen's result about the perfect absorber, the differences in values are caused by different layer thicknesses and different wavelengths [10]. The absorptance monotonously decreases above TIR in both arrangements.

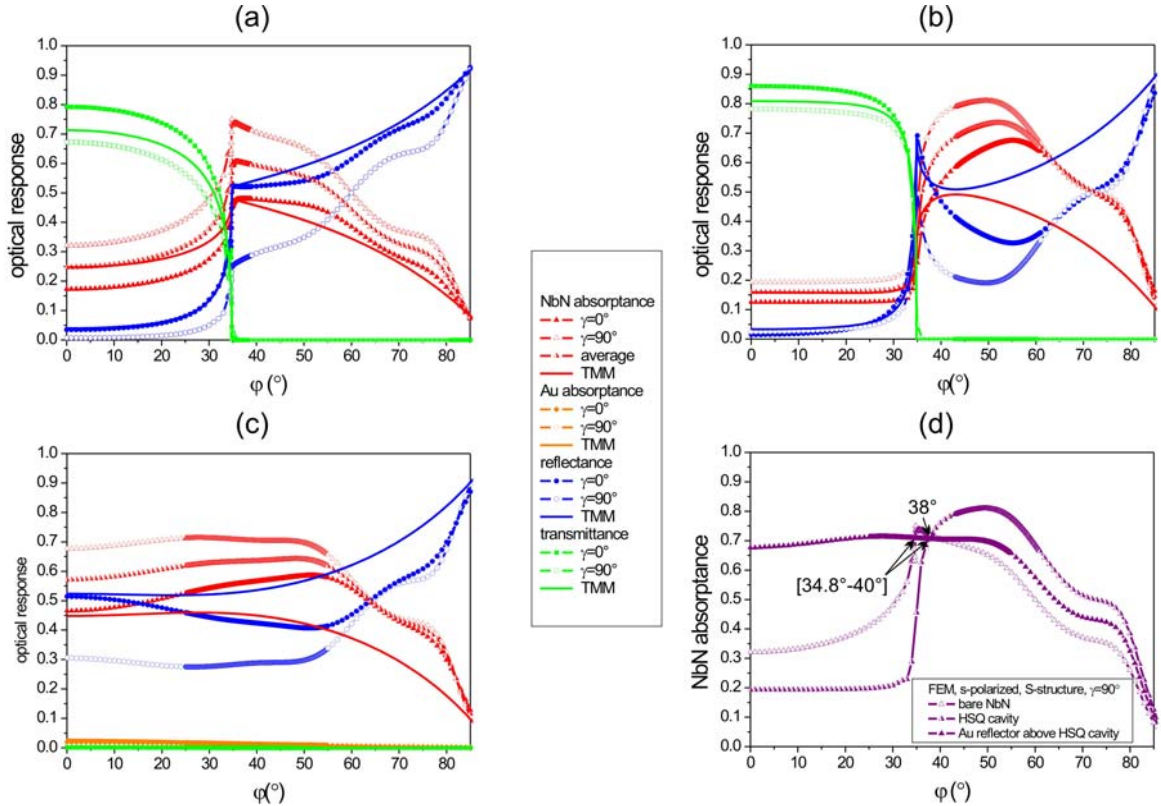

**Fig. 5.** The comparison of optical responses in P-structure configuration ($\gamma$=0°, closed symbols) and in S-structure arrangement ($\gamma$=90°, open symbols) computed by FEM with optical responses determined by TMM (lines) for the case of s-polarized light illumination of NbN patterns (a) standing in air, (b) arrayed below HSQ-filled nano-cavity, (c) aligned below HSQ-filled nano-cavity covered by Au reflector. The results of $\Delta\varphi = 0.05°$ resolution FEM and TMM calculations in (a) [34°, 39°], (b) [43°, 61°], and (c) [25°, 55°] intervals are incorporated into the graphs originating from computation performed with $\Delta\varphi = 1°$ resolution in [0°, 85°] region. (d) Comparison of polar angle dependent absorptance in three optical systems for the case of s-polarized illumination of S-structures.

The transmittance indicates a cut-off at TIR, while the reflectance monotonously increases above TIR. The average of FEM absorptances equals with the TMM absorptance through TIR, while above TIR only the P-structure absorptance approximates the TMM absorptance, but the S-structure absorptance is considerably larger. The stronger absorption appears, when **E**-field oscillation is parallel to NbN stripes similarly to p-polarized illumination of bare NbN patterns.

Important difference is that the **E**-field projection is polar angle independent in s-to-S case, which flattens the optical signals (Fig. 1e). The polar angle independent **E**-field projection is perpendicular to the stripes in s-to-P case (Fig. 1d). This orientation explains the smaller absorptance, and its commensurability with the TMM absorptance, which cannot account for light penetration dependent effects [7-9].

*3.4.2. Optical response of NbN pattern below HSQ-filled nano-cavity*

Figure 5b presents that ATIR phenomenon governs the optical responses for s-polarized light illumination of HSQ-cavity integrated optical system (2).

In case of perpendicular incidence the absorptance is larger in S-structures than in P-structures, and the ratio is reversal in comparison to the p-polarized case. The global absorptance maxima originating from NbN related ATIR appear after inflection points on absorptance curves at TIR, at 49.4° and 55.15°, with values of 80.9 % in S- and 67.3 % in P-structures. These maximal absorptance values are larger than the absorptances observed on bare NbN pattern in analogous arrangements at TIR. The optimal orientation for s-polarized illumination of NbN pattern below HSQ cavity is S-structure arrangement under polar angle corresponding to ATIR.

The transmittance indicates again a cut-off at 34.8° polar angle, while the reflectance curve exhibits a typical ATIR characteristic above TIR caused by presence of lossy NbN stripes below HSQ cavity. The average of the FEM signals equals with TMM absorptance through TIR, but is considerably larger across polar-angles above TIR. The FEM prediction regarding the advantage of orientation under polar angle corresponding to ATIR is in agreement with TMM results, but TMM predicts smaller absorptance, at smaller polar angle.

*3.4.3. Optical response of NbN pattern below Au reflector covered HSQ-filled nano-cavity*

Figure 5c indicates almost unvarying optical responses in contempt of polar angle tuning, when the periodic NbN pattern is illuminated by s-polarized light in gold reflector covered quarter wavelength HSQ cavity.

At perpendicular incidence the absorptance in S-structure arrangement indicates the ~1.5 times ratio reversal with respect to the absorptance observed in case of P-structures [11]. The global maximum appears at 27.85°, where the absorptance is 71.3 % in S-structures. TMM predicts maximum at similar polar angle, but with smaller absorptance value. Interestingly, based on FEM results the global maximum appears in P-structure arrangement at larger 51.5° polar angle, i.e. in the interval corresponding to ATIR, with considerably smaller value of 58.5 %.

The gold absorptance does not indicate any extrema, it monotonously decreases, the transmittance is negligible through the entire polar angle interval, while the reflectance monotonously increases. The average of FEM signals does not equal with the TMM results, FEM predicts considerably larger absorptance in both arrangements through the entire polar angle interval.

*3.4.4. Comparison of different systems illuminated by s-polarized light*

The optimal azimuthal orientation in case of s-polarized illumination is S-structure arrangement, and the available absorptances are compared in Fig. 5d and Table II. The comparison of optical responses observed, when systems (1-3) are illuminated by s-polarized light indicate that it is possible to maximize absorptance by aligning NbN patterns in S-structure arrangement below HSQ nano-cavity, and tilting to polar angle larger than 38°.

| Tilt / Media | Perpendicular | | TIR | | NbN | |
|---|---|---|---|---|---|---|
| | $\varphi$ (°) | A | $\varphi$ (°) | A | $\varphi$ (°) | A |
| air | 0 | 0.32 | 34.85 | **0.75** | - | - |
| HSQ | 0 | 0.193 | - | - | 49.4 | **0.809** |
| Au+HSQ | 0 | 0.676 | 27.85 | **0.71** | 44.7 | 0.703 |

Table II. Summary of absorptance values observed at perpendicular incidence, determined by TIR and NbN-related ATIR phenomenon on polar-angle dependent absorptance of three different optical systems, when the NbN pattern in S-structure arrangement is illuminated by s-polarized light. The absorptances at optimal orientations are indicated in bold.

The absorptance values above this tilting are larger than the NbN absorptance below gold reflector under any polar angle in case of s-polarized light illumination. Theoretically this is the optimal system design, under optimal illumination orientations.

Interestingly, the absorptance is also larger in [34.8°-40°] polar-angle interval, when the NbN pattern in S-structure arrangement is standing in air rather than below Au reflector covered HSQ-filled nano-cavity. The advantage of the reflector is that it results in the most uniformly enhanced absorptance through wide interval of small polar angles.

### 3.5. *Near-field phenomena accompanying polarized light illumination*

The main characteristic in near-field phenomena is that the **E**-field distribution remains symmetrical along the cross-sections of NbN stripes in P-structure-configuration, while it is asymmetrical in S-structure-arrangement (Fig. 6, 7, d-f-to-g-i). This is caused by the different relative orientations of intensity modulations originating from off-axis illumination with respect to NbN pattern. The field enhancement at corners, where the light hits the segments, is maintained as a cost of field depletion along the stripes in S-structures (Fig. 6, 7g-i).

Interestingly, the maximal normalized **E**-field values are larger under analogous polar angles in case of p-polarized illumination of S-structures and during s-polarized illumination of P-structures in all of studied optical systems (Fig. 6, 7d-i). This seems to be in contradiction with the observed smaller absorptances in these arrangements (Fig. 2-5). The explanation is that although large normalized **E**-field is observable at the boundaries and corners outside the NbN segments, when the **E**-field oscillation direction is perpendicular to them, this external enhancement is accompanied by less effective field penetration (Fig. 6g-i, 7d-f). The larger absorptance occurs, when the projection of **E**-field vector is parallel to the stripes, which is known to be due to the larger field penetration [7-9].

The resistive heating values correlate with the absorptance, i.e. the averaged Joule-heating is larger inside the absorbing NbN segments in p-to-P and s-to-S cases (Fig. 6, 7j-l, red–to–blue curves).

Interestingly, the characteristic of the resistive heating cross-section only slightly varies with the polar angle. Although the maximal values are similar, the resistive heating indicates significantly different distribution in the two different azimuthal orientations. The highest averaged values are observable at orientations corresponding to global maxima on the absorptance in both arrangements.

*3.5.1. Near-field phenomena accompanying p-polarized light illumination*

The most intensive **E**-field penetrating into bare NbN stripes is observable at global maxima related to ATIR, due to evanescent-field concentration at substrate-NbN interface (Figure 6d, g). The more homogeneous field distribution surrounding the NbN segments ensures to reach the highest absorptance in P-structure configuration at 48° polar angle (Fig. 6d). The comparison of line cross-sections taken at the middle of NbN stripes proves that the most effective penetration is observable at the global maxima originated from ATIR (Figure 6j).

Figure 6e, h provides the near-field explanation for the optical response observed at the polar angle corresponding to TIR at sapphire-air interface. Due to the phase shift introduced by quarter wavelength HSQ cavity, the incoming and reflected fields are in phase around the NbN stripes. As a result a significant part of the EM-field might be absorbed in the lossy film, similarly to s-polarized illumination of bare NbN patterns [10]. This explains the appearance of the global maximum on NbN absorptance curves. The line cross-sections indicate commensurate resistive heating maxima, but the NbN segments are more uniformly heated in P-structure configuration (Fig. 6k).

Figure 6f, i indicate that at perpendicular incidence there is an **E**-field antinode at the bottom of HSQ-cavity due to the gold reflector, which explains the very large available absorptance. As a result, perpendicular incidence is the optimal illumination of this system in both arrangements.

The line cross-sections taken at the middle of the NbN stripes indicate that the largest Joule-heating values are observable at perpendicular incidence according to the absolute maxima on the absorptance (Figure 6l).

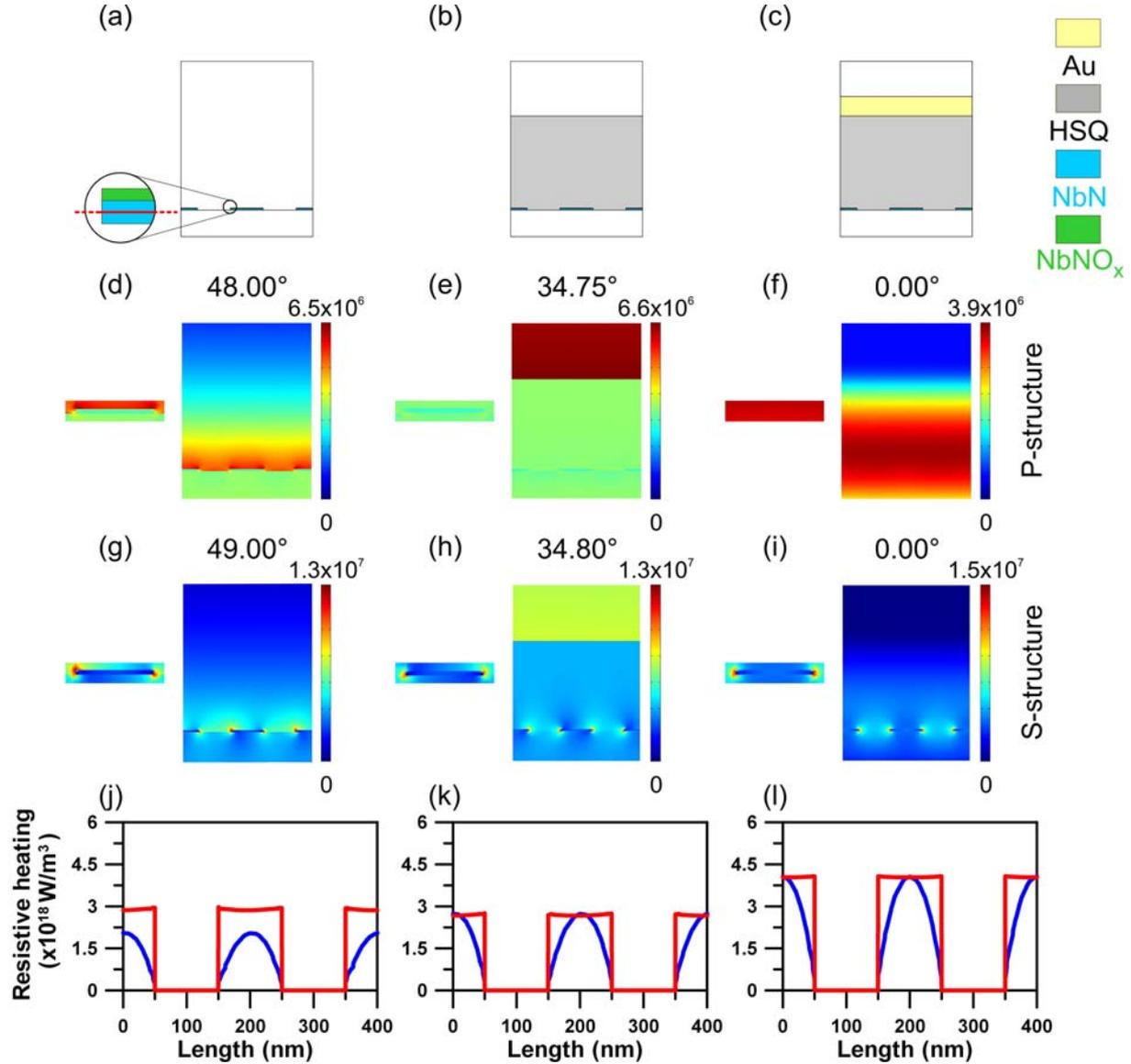

**Fig. 6.** (a-c) Schematic drawing showing two unit cells in a plane perpendicular to the 200 nm periodic NbN pattern, where the near-field cross-sections are investigated on NbN pattern (a, d, g, j) standing in air, (b, e, h, k) below HSQ nano-cavity, and (c, f, i, l) below Au reflector covered HSQ nano-cavity. The normalized **E**-field distribution due to p-polarized light illumination at the extrema observed in (d-f) P-structure configuration, and in (g-i) S-structure arrangement. The cross-sections indicate at the global maxima on NbN absorptance (d, g) evanescent field enhancement due to ATIR phenomenon, (e, h) phase-matched fields due to HSQ cavity and (f, i) **E**-field antinode due to gold reflector. These pictures are presented on a scale to illustrate better small variations in the field, and the insets indicate the **E**-field distribution locally around single NbN stripes. (j-l) The comparison of the resistive heating at the extrema inside NbN segments in P-structure configuration (red) and in S-structure arrangement (blue) along the horizontal line indicated on the (a) inset.

*3.5.2. Near-field phenomena accompanying s-polarized light illumination*

On s-polarized light illuminated NbN pattern standing in air, large **E**-field with continual phase transition through NbN stripes is observable at TIR corresponding to the global absorptance maximum (Fig. 7d, g).

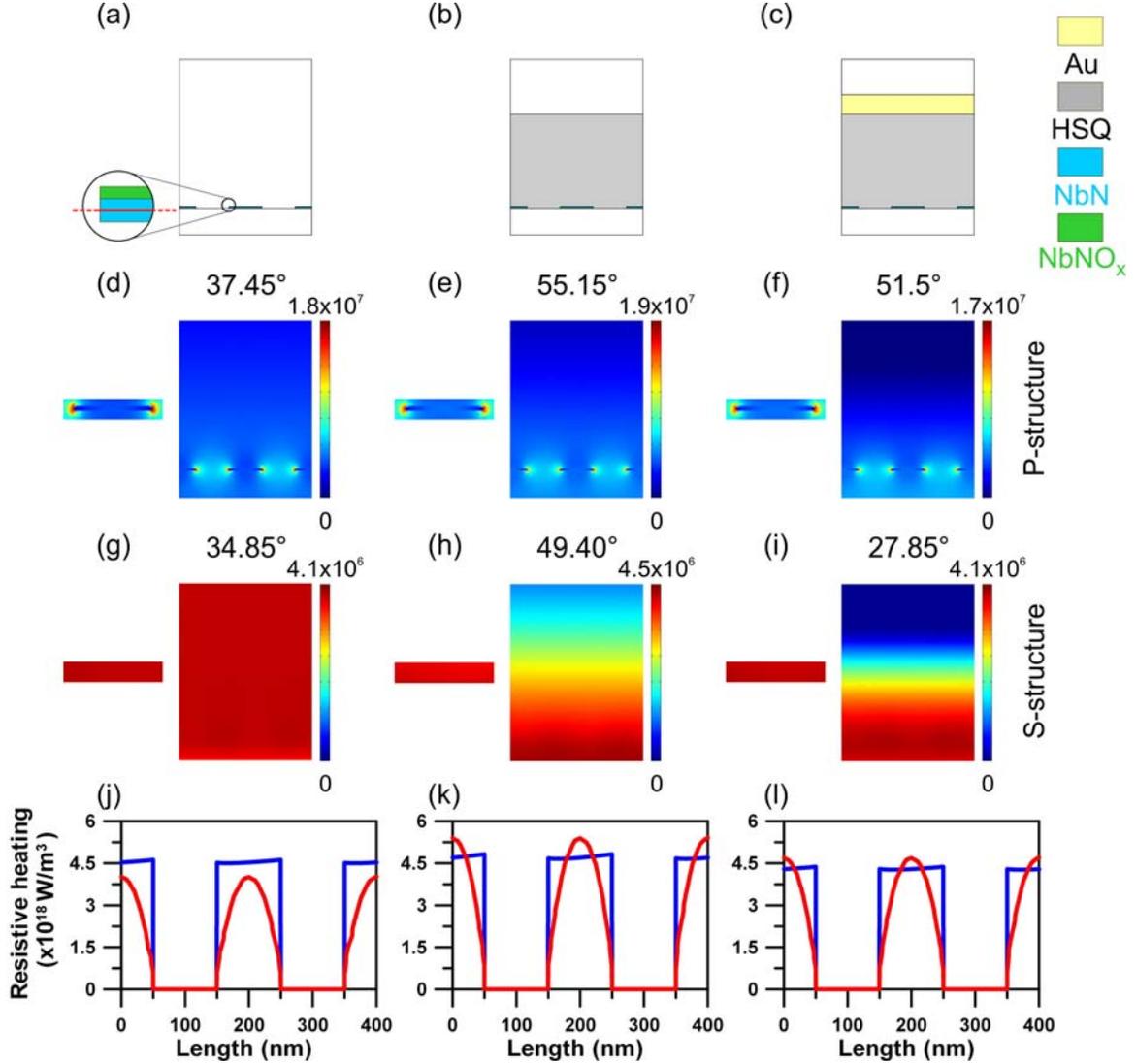

**Fig. 7** (a-c) Schematic drawing showing two unit cells in a plane perpendicular to the 200 nm periodic NbN pattern, where the near-field cross-sections are investigated on NbN pattern (a, d, g, j) standing in air, (b, e, h, k) below HSQ nano-cavity, (c, f, i, l) below Au reflector covered HSQ nano-cavity. The normalized **E**-field distribution due to s-polarized light illumination at the extrema observed in (d-f) P-structure configuration, and in (g-i) S-structure arrangement. The cross-sections indicate at the global maxima on NbN absorptance (d, g) phase-matched fields due to s-polarized illumination, (e, h) enhanced evanescent **E**-field concentration below HSQ cavity (f, i) **E**-field antinode due to gold reflector. These pictures are presented on a scale to illustrate better small variations in the field, and the insets indicate the **E**-field distribution locally around NbN stripes. (j-l) The comparison of the resistive heating at the extrema inside NbN segments in P-structure configuration (red) and in S-structure arrangement (blue) along the horizontal line indicated on the inset in (a).

The cross-section about the resistive heating indicates that larger and more uniform heating occurs in S-structure arrangement (Fig. 7j).

At the bottom of HSQ cavity an evanescent **E**-field concentration is observable at substrate-NbN interface at polar angle corresponding to ATIR-related absorptance maximum, which is further enhanced due to the phase-shift introduced by the cavity (Fig. 7e, h). Even though the maximal resistive heating is larger in P-structures, the heating is more compensated in S-structures (Fig. 7k).

At both global maxima, which appear in different polar angle intervals, there is an **E**-field anti-node at the bottom of the cavity due to the HSQ-cavity covered by gold reflector (Fig. 7f, i). The resistive heating exhibits same relation as below HSQ cavity, with slightly reduced amplitudes (Fig. 7l).

## 5. Conclusion

The most important result of these calculations is that optimization of the illumination angle of absorbing NbN patterns can help to optimize the SNSPDs absorptance, as a result all of three studied optical systems are promising in specific application areas. While the course of dual-angle dependent variations fundamentally differs in the three SNSPD device designs in case of specific polarizations, each system design has its own optimal illumination condition. Based on FEM computation results it was possible to select the optimal relative orientation of the NbN stripes with respect to the incidence plane of the polarized light. This is the P-structure configuration in case of p-polarized illumination, and S-structure arrangement during s-polarized light illumination, both correspond to **E**-field oscillation parallel to the stripes. The optimal polar angles resulting in maximal absorptance were determined too.

When p-polarized light is applied for illumination, these are the polar angle resulting in NbN-related ATIR in case of bare NbN pattern, smaller polar angle corresponding to TIR at sapphire-air interface in case of NbN stripes below HSQ nano-cavity, and perpendicular incidence in an optical system consisting of gold reflector covered HSQ nano-cavity. For s-polarization the global maximum is at TIR in case of bare NbN pattern, which is shifted to region of ATIR below HSQ nano-cavity, and compensated through a wide polar angle interval below gold reflector covered HSQ nano-cavity.

**Acknowledgement**

This work has been supported by the U.S. Dept. of Energy Frontier Research Centers program and by the Hungarian OTKA foundation from the National Innovation Office (NIH), under grants No OTKA-NKTH CNK 78459 and OTKA-NKTH K 75149. Maria Csete thanks the Balassi Institute for the Hungarian Eötvös post-doctoral fellowship. The authors thank for the helpful discussions with E. Driessen and M. de Dood and Comsol engineers in Burlington. Professor Karl K. Berggren contributed to the initial concept of the paper by suggesting the comparison of different optical systems, Áron Sipos and Faraz Najafi prepared the models for numerical simulation, while Mária Csete analyzed the results and prepared the manuscript.

**References**

[1] G. N. Gol'tsman, O. Okunev, G. Chulkova, A. Lipatov, A. Semenov, K. Smirnov, B. M. Voronov, A. Dzardanov, C. Williams and R. Sobolewski, „Picosecond superconducting single-photon optical detector," *Appl. Phys. Lett.* **79**(6) 705-708 (2001).

[2] F. Marsili, F. Najafi, E. Dauler, F. Bellei, X. Hu, M. Csete, R. Molnar and K. K. Berggren: „Single-photon detectors based on ultranarrow superconducting nanowires," *Nano Letters* **11**(5) 2048-2053 (2011).

[3] K. M. Rosfjord, J. K. W. Yang, E. A. Dauler, A. J. Kerman, V. Anant, B. M. Voronov, G. N. Gol'tsman and K. K. Berggren, "Nanowire Single-Photon detector with an integrated optical cavity and anti-reflection coating," *Optics Express* **14**(2) 527-534 (2006).

[4] S. Miki, M. Takeda, M. Fujiwara, M. Sasaki, Z. Wang: "Compactly packaged superconducting nanowire single-photon detector with an optical cavity for multichannel sysrem," *Optics Express* **17**(226) 23557-235640 (2006).

[5] X. Hu, C. W. Holzwarth, D. Masciarelli, E. A. Dauler and K. K. Berggren: Efficiently coupling light to superconducting nanowire single-photon detectors," *IEEE Trans. Appl. Supercond.* **19**, 336-340 (2009).

[6] X. Hu, E. A. Dauler, R. J. Molnar, K. K. Berggren, „Superconducting nanowire single-photon detectors integrated with optical nano-antennae," *Optics Express* **19**(1) 17-31 (2011).

[7] E. A. Lewis, J. P. Casey: "Electromagnetic reflection and transmission by gratings of resistive wires," *J. Appl. Phys.* **23** 605-608 (1952).

[8] E. F. C. Driessen, F. R. Braakman, E. M. Reiger, S. N. Dorenbos, V. Zviller, M. J. A. de Dood, „Impedance model for the polarization-dependent optical absorption of superconducting single-photon detectors," *The European Journal Applied Physics* **47**(1071) 1-6 (2009).

[9] V. Anant, A. J. Kermann, E. A. Dauler, J. K. W. Yang, K. M. Rosfjord, K. K. Berggren, „Optical properties of superconducting nanowire single-photon detectors," *Optics Express* **16**(14) 10750-10761 (2008).

[10] E. F. C. Driessen and M. J. A. de Dood, "The perfect absorber," *Appl. Phys. Lett.* **94**(171109) 1-3 (2009).

[11] M. Csete, Á. Sipos, F. Najafi, X. Hu, K. K. Berggren, "Numerical method to optimize the polar-azimuthal orientation of infrared superconducting nanowire single photon detectors," *Appl. Opt.* **50**(31) 5949-5956 (2011).

[12] M. Csete, A. Szalai, F. Najafi, K. K. Berggren, "Impact of polar-azimuthal illumination angles on nano-cavity-array integrated superconducting nanowire single-photon detectors efficiency" publication in progress.

[13] M. Born and E. Wolf, *Principles of Optics*, Pergamon Press, Cambridge (1964).

[14] E. Popov, S. Enoch, G. Tayeb, M. Neviére, B. Gralak, N. Bonod: "Enhanced transmission due to nonplasmon resonances in one- and two-dimensional gratings," *Appl. Opt.* **43**(5), 999-1008 (2004).